\documentclass[journal]{IEEEtran}
\usepackage{subfigure}
\usepackage{graphicx} % 
\usepackage{amsmath,amssymb}
\usepackage{bbm}
\usepackage{xcolor}
\usepackage{stfloats}
\usepackage{setspace}
\usepackage{alltt, verbatim}
\title{Analog Beamforming Aided by Full-Dimension One-Bit Chains}

\author{Lina Liu$^*$, Weimin Xiao$^\dagger$, Jialing Liu$^\dagger$, and Zhigang Rong$^\dagger$\thanks{
This content originates from the research conducted during an internship at Futurewei Technologies.}\\
$^*$Department of Electrical and Computer Engineering, Northwestern University\\
$^\dagger$Wireless Standard\re{s} and Research, Futurewei Technologies
}
\date{\today}

\definecolor {ForestGreen} {RGB} {34,139,34}

\newcommand{\re}[1]{{\color{black}#1}}
\newcommand{\dg}[1]{{\color{black}#1}}

\newcommand{\ba}{{\mathbf{a}}}
\newcommand{\bb}{{\mathbf{b}}}
\newcommand{\bee}{{\mathbf{e}}}
\newcommand{\br}{{\mathbf{r}}}
\newcommand{\bs}{{\mathbf{s}}}
\newcommand{\bw}{{\mathbf{w}}}
\newcommand{\bx}{{\mathbf{x}}}
\newcommand{\bz}{{\mathbf{z}}}

\newcommand{\bH}{{\mathbf{H}}}
\newcommand{\bI}{{\mathbf{I}}}
\newcommand{\bR}{{\mathbf{R}}}
\newcommand{\bS}{{\mathbf{S}}}
\newcommand{\bX}{{\mathbf{X}}}
\newcommand{\bZ}{{\mathbf{Z}}}

\newcommand{\bbC}{{\mathbb{C}}}
\newcommand{\bbE}{{\mathbb{E}}}

\newcommand{\utheta}{{\underline{\theta}}}

\newcommand{\calQ}{{\mathcal{Q}}}
\newcommand{\calZ}{{\mathcal{Z}}}
\newcommand{\calCN}{{\mathcal{CN}}}

\newcommand{\sign}{{\text{sign}}}

\newcommand{\real}{{\text{Re}}}
\newcommand{\imag}{{\text{Im}}}

\DeclareMathOperator*{\newargmax}{arg\,max}

\graphicspath{{fig/}{../github_fig/}}

\begin{document}
	
\maketitle

\begin{abstract}
This paper investigates the design of analog beamforming at the receiver in millimeter-wave (mmWave) multiple-input multiple-output (MIMO) systems, aided by full digital chains featuring 1-bit \re{analog-to-digital converters} (ADCs). We advocate utilizing full digital chains to facilitate rapid beam acquisition for subsequent communication, even without prior knowledge of the training pilots. To balance energy consumption and implementation costs, we opt for 1-bit ADCs. 
We propose a two-step maximum likelihood (ML)-based algorithm to estimate angles of arrival (AoAs) and \re{provide} the design of analog beamforming to maximize the received signal-to-noise ratio (SNR). \re{For narrowband coherent channels, we estimate multiple AoAs and propose a beamforming approach that incorporates all estimated AoAs. For wideband channels, we propose beamforming towards the direction that captures significant energy from all clusters. The effectiveness of the narrowband beamforming scheme is validated through synthetic tests, while the wideband beamforming scheme is evaluated under the 3GPP clustered-delay-line (CDL)-C channel model.}
\end{abstract}

\section{Introduction}
With its extensive bandwidth, millimeter-wave (mmWave) communication holds the potential to deliver high data rates, low latency, and high capacity, especially when combined with multiple-input multiple-output
(MIMO) technology. However, realizing these benefits at reasonable costs has proven challenging \cite{GSMAwhitepaper}. 
A key issue is the rapid mmWave channel variations in mobile scenarios, requiring timely channel estimation for reliable communication. Moreover, deploying digital chains for numerous antenna elements is expensive and energy-intensive, as is the use of high-resolution analog-to-digital converters (ADCs) \cite{el2014spatially}.
A common solution to the latter challenge is a hybrid architecture with fewer digital chains than transmit/receive antennas \cite{zhang2019bussgang}, but this can prolong channel estimation \re{due to the need for beam sweeping}, hindering the beamformer optimization.
In this paper, we consider leveraging received signals from a full set of digital chains with 1-bit ADCs to compute an analog beamformer for subsequent data communication. This approach aims to expedite 
beam acquisition at the moderate cost of digital chains with 1-bit ADCs \re{without the need for beam sweeping}. 
Since these digital chains operate \re{only when beam acquisition is needed}, their energy consumption can be kept minimal. 

Several studies have tackled channel estimation in mmWave MIMO systems with low-resolution ADCs. Methods include formulating quantized compressed sensing problems \cite{mo2018channel,xu2020angular}, using Bussgang decomposition to linearize quantization \cite{zhang2019bussgang,srinivas2022channel}, incorporating quantization constraints into optimization \cite{esfandiari2023admm}, and integrating quantization into likelihood-based objectives \cite{liu2020angular,nguyen2023deep}. By leveraging channel sparsity in the angular domain, the estimation problem is often treated as sparse support recovery \cite{mo2018channel,xu2020angular,zhang2019bussgang,srinivas2022channel}. Optimization and machine learning tools are also applied depending on the formulation \cite{esfandiari2023admm,nguyen2023deep}. In narrowband coherent channels with uniform linear arrays (ULAs), a maximum likelihood (ML)-based approach estimates angles of departure/arrival (AoDs/AoAs) and path coefficients iteratively, progressively refining estimates path by path until convergence \cite{liu2020angular}.

In mmWave communications, analog beamforming reduces energy consumption by processing multi-antenna operations in the analog domain, addressing the energy-intensive nature of ADCs \cite{abbas2017millimeter}. Without direct channel estimation, codebook-based beamforming schemes use hierarchical codebooks to enhance efficiency \cite{he2015suboptimal,sun2017analog}. A sectional search strategy \re{performs beam sweeping to} identify the codewords that best align with the channel's AoD/AoA. 
However, each beamforming vector test requires a measurement, \re{leading to high overhead and latency but limited beam resolution}. In contrast, full digital chains \re{enable one-shot estimation with a moderate number of measurements, allowing high-resolution beamforming through signal processing even when using low-resolution ADCs.}

% In \cite{sun2017analog}, joint channel training and analog beamforming optimization are achieved through multi-sectional search of a designed two-step codebook. In \cite{fonteneau2022efficient}, analog beamforming in wideband MIMO-orthogonal frequency division multiplexing (OFDM) systems is explored, involving the evaluation of beamforming on each subcarrier, followed by eigen-decomposition-based beamforming across all subcarriers. 
% and applied in various contexts including multiuser detection \cite{choi2015analog}, full-duplex communication \cite{lopez2019analog}, and downlink multiuser transmission \cite{jiang2018multi}.

In channel estimation literature, the goal is typically to minimize the error in estimating the channel matrix, often quantified by the normalized mean squared error (NMSE). 
However, NMSE fails to directly reflect received signal-to-noise ratios (SNRs) and treats all channel coefficients equally, disregarding their varying influence on beamforming. \re{We leverage estimated angular information to design analog receive beamforming, aiming to maximize the post-beamforming SNR.}
Notably, spatial multiplexing is primarily influenced by the angular-domain information of the channel, which remains unaffected by time-domain factors such as channel coherence (specifically path coefficients) and varying transmit pilots.
Our contributions are summarized as follows:
\begin{itemize}
\item In the special case of a single AoA and zero delay spread, we have developed a principled, ML-based method for AoA estimation, regardless of channel coherence or \re{availability of} pilot knowledge. 
\item We extend the method to handle scenarios with multiple AoAs and zero delay spread. Unlike \cite{liu2020angular}, we extract essential beamforming information: AoA and \re{effective} path gain, which accounts for the transmit antenna response, transmitted pilot, and path coefficient. Our method employs a one-shot coarse estimation followed by fine estimation for each AoA, applicable to both ULAs and UPAs.
Accurate angle estimates are achieved when paths are sufficiently separated, allowing the proposed beamformer to recover over 90\% of the received power, \re{compared to ideal} beamforming with full channel knowledge.
\item  We adapt the method to handle the scenario of multiple clusters of AoAs with delay spread. Testing is conducted using signals generated based on the 3GPP's clustered-delay-line (CDL)-C channel model \cite{ETSI-TR}. A tailored beamforming approach is proposed to identify a beam direction that captures significant energy from all clusters. We evaluate the effectiveness of this approach by comparing it to the optimal beamformer under ideal conditions, assuming unquantized signals and no additive noise.
\end{itemize}

\textit{Notations:} Let $a$ denote a scalar, $\mathbf{a}$ a column vector, $\mathbf{A}$ a matrix, and $\mathcal{A}$ a set. $(\cdot)^T$ and $(\cdot)^H$ represent the transpose and conjugate transpose, while $\lvert\cdot\rvert$ indicate the absolute value. $\mathbf{0}_{M}$ and $\mathbf{I}_{M}$ refer to the $M$-dimensional zero vector and the $M \times M$ identity matrix. Additionally, $\text{sign}(\cdot)$, $\text{Re}(\cdot)$, and $\text{Im}(\cdot)$ denote the sign, real part, and imaginary part, respectively.

\section{System Model and Problem Description}
\label{s:model}
\begin{figure}[htbp]
\centering
\includegraphics[width=0.3\textwidth]{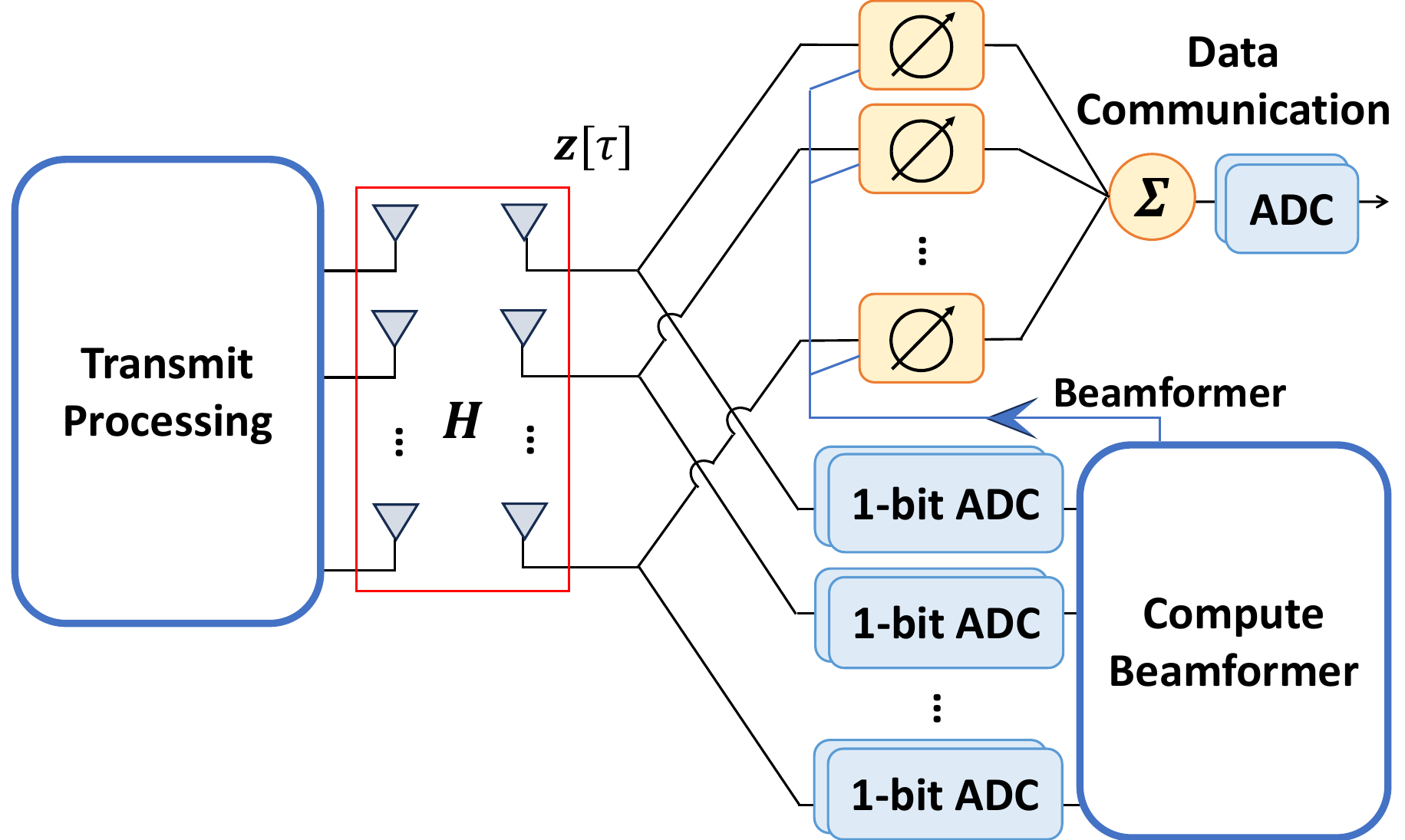}
\caption{The receive processing involves two \re{stages} using the same receiving antennas: one for beamforming design, and the other for data communication.}
\label{fig:receive_processing}
\end{figure}

We consider an mmWave MIMO system with $M_t$ transmit antennas and $M_r$ receive antennas. We focus on the receiver, assuming effective beamforming at the transmitter. Each communication period within the channel coherence time has two \re{stages}. In the first \re{stage}, we utilize a full set of digital \re{receive} chains equipped with 1-bit ADCs to estimate an analog beamformer.
In the second \re{stage}, the computed analog beamformer is employed for data communication using a single digital chain with a high-resolution ADC. 
These two \re{stages} repeat in subsequent periods, \re{with the first stage executed only during initial beam acquisition or beam recovery}. A block diagram of the system is illustrated in Fig.~\ref{fig:receive_processing}.

\subsection{The mmWave MIMO Channel}
Define $\ba_{M_r}$ and $\ba_{M_t}$ as the antenna array response vectors at the receiver and transmitter, respectively. \re{Assume the mmWave channel consists of $L$ propagation paths. 
Let $\alpha_{l}$, $\delta_{l}$, $\varphi_{l}$,
$\theta_{l}$,
$\omega_{l}$, and $\psi_{l}$ represent the complex coefficient, delay, azimuth and \re{elevation} angles of arrival, and azimuth and \re{elevation} angles of departure of the $l$-th path, respectively.} Let $T$ denote the \re{chip} duration and $p(\cdot)$ represent the waveform that captures pulse shaping and analog/digital filtering effects. Assuming a delay spread of $D$ \re{chip} intervals, the baseband channel impulse response at lag $d$ is given by \cite{mo2018channel}
\begin{align} \label{eq:Cluster_channel}
  \bH[d] &=\sum_{l=1}^{L} \alpha_{l} \ba_{M_r}(\varphi_{l}, \theta_{l}) \ba_{M_t}^{H}(\omega_{l},\psi_{l}) p( dT - \delta_{l}).
\end{align}
% Paths can be grouped into clusters, with distinct departure and arrival angles between clusters and minimal deviations within each cluster.
At time $\tau$, the received signal \re{without} noise is expressed as
\begin{align}\label{eq:sig_comp}
\bx[\tau]=\sum_{d=0}^{D-1}\bH[d]\bs[\tau-d], 
\end{align}
where $\bs[\tau] \in \mathbb{C}^{M_t\times 1}$ denotes the transmitted symbol at time $\tau$.
% denote  during data communication. 
% Let
% \begin{align}
%     \bz[\tau] = \bx[\tau] + \bw[\tau],
% \end{align}
% where $\bw[\tau] {\sim} \mathcal{CN}(\mathbf{0}_{M_r}, \bI_{M_r})$ represents the additive white Gaussian noise.
Let $\bw[\tau] \sim \mathcal{CN}(\mathbf{0}_{M_r}, \bI_{M_r})$ denote the additive white Gaussian noise. The unquantized received signal is given by 
\begin{align} 
\bz[\tau] = \bx[\tau] + \bw[\tau]. 
\end{align}

\subsection{The Beamformer Estimation \re{Stage}}
Following each receive antenna, %employs 
a digital chain is equipped with a pair of 1-bit ADCs which % totaling $2M_r$ ADCs that 
separately quantize the real and imaginary parts of the received signal. % from each antenna. 
Let the quantization function be described as $r =\calQ(z) =  \sign \left( \real(z) \right) + j  \sign \left( \imag(z) \right)$,
where $j $ represents the imaginary unit.
% , and $\sign(\cdot),\real(\cdot)$ and $\imag(\cdot)$ represent taking the sign, real and imaginary parts of their arguments, respectively. 
 The quantized output $\br[\tau]\in \mathbb{C}^{M_r \times 1}$ at time $\tau$ (in the estimation \re{stage}) can be written as
\begin{equation} \label{eq:r[tau]}
\br[\tau] =  \calQ \left(
\bz[\tau]
%\bz_{\text{sig}}[\tau]
%\sum_{d=0}^{D-1} \bH[d] \bs[\tau-d] 
%+ \bw[\tau] 
\right),
\end{equation}
where the quantization function is applied element-wise to the vector argument.

% The objective is to estimate the channel information using the signals described in~\eqref{eq:r[tau]} to facilitate beamforming design.

\subsection{The Analog Beamforming \re{Stage}}
%for Data Communication}
\label{s:problem}
Throughout this paper, analog beamformers are implemented using phase shifters. A single digital chain with \re{$M_r$ phase shifters and} a full-resolution ADC is used at the receiver during data communication \re{stage}, where the quantization error is neglected. %\dg{Here we neglect the quantization error by assuming that it is dominated by some additive noise.}
Let $\xi_m$ denote the phase of the $m$-th phase shifter. The analog beamformer
%Denote by 
$\mathbf{b} \in \mathbb{C}^{M_r \times 1}$ % \dg{denote} the analog beamformer, whose $m$-th element represents the phase shift applied to the received %(unquantized) 
% signal at the $m$-th receive antenna, i.e., 
is determined by $b_m=e^{j \xi_m}$.
% with some designed phase $\xi_m$. 
At time $\tau$ (in the beamforming \re{stage}), the signal after beamforming can be expressed as
\begin{align}
y[\tau]=\bb^{{H}} \bz[\tau] .
% \bz_{\text{sig}}[\tau]+\bb^{{H}}\bw[\tau].
\end{align} We aim to find the optimal solution to the following problem
\begin{align}\label{eq:beamforming_prob}
\max_{\bb:\,
|b_1|=\dots=|b_{M_r}|=1
% \vert b_m\vert=1,\forall m=1,\cdots,M_r
}\quad{\bbE_{\tau}\left[\left\vert\bb^{{H}}\bx[\tau]\right\vert^2\right]}.
\end{align}
% because $\bbE_{\tau}\left[\left\vert\bb^{{H}}\bw[\tau]\right\vert^2\right]=M_r$.

\section{Angular-Domain Channel Estimation}
\label{sec:method}
To address the channel estimation problem, we begin by considering the simplified model with zero delay spread, which yields the following channel description
\begin{align}\label{eq:flat_channel}
  \bH = \sum_{l=1}^{L}\alpha_{l} \ba_{M_r}(\varphi_{l}, \theta_{l}) \ba_{M_t}^{H}(\omega_{l},\psi_{l}).
\end{align}
% The quantized received signal becomes
% \begin{equation} \label{eq:r}
% \br[\tau] =  \calQ \left(\bH\bs[\tau] + \bw[\tau] \right).
% \end{equation}
We differentiate the channel models by designating~\eqref{eq:Cluster_channel} as the wideband coherent channel and~\eqref{eq:flat_channel} as the narrowband coherent channel. First, we present an angular-domain channel estimation method based on a simplified narrowband model. We then extend the algorithm to the wideband model.

\subsection{ML-based Angle Estimation for Narrowband Coherent Channel with ULA and Repetitive Pilot Symbols}
\label{sec:coherent_ULA}
%Our illustration begins 
\dg{We begin} with the scenario where the antenna is configured as a ULA at both the transmit and receive sides. 
% In this setup, the antenna response can be characterized by the arrival/departure angle and array size. 
Let $\theta\in\left[-\frac{\pi}{2},\frac{\pi}{2}\right]$ denote the AoA at the receiver. Let the antenna spacing be half of the wavelength. With %$M_r$ receive 
$M$ antennas, the antenna response vector can be represented as 
\begin{align}\label{eq:ULA_reponse}
\ba_M(\theta) = \left[1,e^{j \pi\sin(\theta)},\cdots,e^{j \pi\sin(\theta)(M-1)}\right]^T. %,
% \ba_{M_r}(\theta) = \left[1,e^{j \pi\sin(\theta)},\cdots,e^{j \pi\sin(\theta)(M_r-1)}\right]. %,
\end{align}
% Let $\ba_{M_t}(\cdot)$ be defined the same way as in~\eqref{eq:ULA_reponse} with $M_r$ replaced by $M_t$.
% where the subscript $r$ pertains to the receiver with $M_r$ antennas. 
The channel matrix can be further simplified as $\bH = \sum_{l=1}^{L} \alpha_{l} \ba_{M_r}(\theta_{l}) \ba_{M_t}^{H}(\psi_{l})$,
where $\theta_l$ and $\varphi_l$ denote the AoA and AoD of path $l$, respectively. 
Let the transmitted signal $\bS$ be a repetition of an unknown \re{signal} $\bs$, i.e., $\bS = [\bs, \dots, \bs] \in \bbC^{M_t \times N_d}$, where $N_d$ is the length \re{of the unknown signal}.

% \begin{figure}
% \subfigure[ULA coordinate]{
% \begin{minipage}{0.225\textwidth}
%     \centering
%     \includegraphics[width=0.45\textwidth]{ULA_coor}
% \end{minipage}
% \label{fig:ULA_coor}
% }
% \subfigure[UPA coordinate]{
% \begin{minipage}{0.225\textwidth}
%     \centering
%     \includegraphics[width=0.75\textwidth]{UPA_coor}
% \end{minipage}
%     \label{fig:UPA_coor}
% }
% \caption{Coordinate definition for receive antenna response. The same coordinate system applies to the transmitter side.}
% \end{figure}

\subsubsection{A Single Path}\label{sec:ML_ULA_single_path}
In the special case of a single path, we neglect the subscript $l$. 
The received signal at the $m$-th receive antenna in time slot $\tau$ before quantization can be expressed as
\begin{align}
z_m[\tau] = \zeta e^{j \pi(m-1)\sin(\theta)} + w_m[\tau],
\end{align}
where $\zeta = \alpha \mathbf{a}_{M_t}^{H}(\psi) \mathbf{s}$ 
represents the effective path gain.
\re{Throughout this paper, we use $|\zeta|^2$ to characterize the path SNR.}
To derive the likelihood of observing the output $\bR = \left[\br[1], \br[2], \cdots, \br[N_d]\right]$ given $\theta$, we assume $\zeta$ follows a uniform distribution over $\calZ = \left\{ \gamma e^{j  2\pi\frac{k}{N_{\zeta}}} \right\}_{k=0}^{N_{\zeta}-1}$, where $\gamma$ denotes the amplitude, and $N_{\zeta}$ represents the number of discrete phases\footnote{\re{The peak locations of the likelihood function are insensitive to the assumed distribution. For implementation, we assume the uniform distribution with $N_{\zeta} = 100$ and $\gamma = 0.1$, corresponding to a -20 dB path SNR, for simplicity.}}.  Before we present the likelihood expression, let us first introduce the following function:
\begin{align}
f_N(\upsilon,\lambda) \equiv Q\left(-\sqrt{2}\upsilon\right)^{\lambda}\left(1-Q\left(-\sqrt{2}\upsilon\right)\right)^{N-\lambda}.
    % f_N(\upsilon,\lambda) = 2^N Q(\sqrt{2}\upsilon))^{\lambda}(1-Q(\sqrt{2}\upsilon))^{N-\lambda}.\\
    % f_N(\upsilon,\lambda) = (\text{erfc}(\upsilon))^{\lambda}(1+\text{erf}(\upsilon))^{N-\lambda}
\end{align}
% In this function,
where $Q(a)$ denotes the Q-function.
% of a Gaussian random variable with mean equal to $\upsilon$ and variance equal to $1/2$ is negative.
% $\mathcal{N}\left(-\upsilon,\frac{1}{2}\right)$ exceeding $0$  (discounting a constant $1/2$). 
% Conversely, $(1+\text{erf}(\upsilon))$ signifies the probability of such a variable falling below $0$ (discounting a constant $1/2$). 
% We interpret $f_N(\upsilon,\lambda)$ as the probability of observing $N$ measurements of a real constant $\upsilon$, corrupted by independent zero-mean Gaussian noise with variance $1/2$, where  $\lambda$ of these measurements are quantized to $1$ by a 1-bit ADC.
\re{We interpret $f_N(\upsilon,\lambda)$ as the probability of observing $N$ given measurements of a real constant $\upsilon$, after corrupted by independent zero-mean Gaussian noise with variance $1/2$ and passing through a 1-bit ADC, where $\lambda$ of these measurements are quantized to $1$.}
We further define
\begin{align}
&\rho_{m}(\theta,\zeta)\equiv\real\left(\zeta e^{j \pi(m-1)\sin(\theta)}\right),\\
&\kappa_{m}(\theta,\zeta)\equiv\imag\left(\zeta e^{j \pi(m-1)\sin(\theta)}\right).
\end{align}
Neglecting a constant multiplier, the likelihood function can be represented as 
\begin{align}\label{eq:obj_coh_ULA}
g(\theta)= % \nonumber\\
&\sum_{\zeta\in\calZ}\prod_{m=1}^{M_r}
f_{N_d}(\rho_m(\theta,\zeta), \mu_{m})
f_{N_d}(\kappa_{m}(\theta,\zeta), \nu_{m}) ,
\end{align}
where
$\mu_{m} = \sum_{\tau=1}^{N_d}\big(\real(r_m[\tau])+1\big)/2$, $\nu_m = \sum_{\tau=1}^{N_d}\big(\imag(r_m[\tau])+1\big)/2$, \re{ and $r_m[\tau]$ is the $m$-th element of $\br[\tau]$ as defined in Eq. \eqref{eq:r[tau]}}. \dg{(\re{Numerically}, using the log-likelihood helps to keep the precision.)}
% Certain irrelevant constants are omitted as they do not impact the optimization problem. Moreover, in implementation, 
% taking the logarithm of the objective helps manage scales effectively.
The objective function defined in~\eqref{eq:obj_coh_ULA} is nonconvex, but it exhibits a distinct and narrow peak around the true AoA.
% \dg{if beamforming can boost the SNR}. 
Leveraging this observation, we propose a two-step estimation approach for $\theta$. Initially, we sample the AoA according to
\begin{align}\label{eq:crude_sample}
\utheta_q=\arcsin\left(-1+(q-1)/{M_r}\right), \quad 1\leq q\leq 2M_r,
\end{align} 
and compute corresponding objective function values. The most prominent sampling peak provides a coarse estimate, and a constrained region around it is used for fine estimation with a gradient-based algorithm.
Suppose $\check{q}$ \dg{maximizes $g\left(\utheta_q\right)$ for $q\in\{1,\dots,2M_r\}$.}
% =\newargmax_q\left\{g\left(\utheta_q\right)\right\}_{q=1}^{2M_r}$. 
The constrained feasible region is set as $\left[\utheta_{\check{q}-1},\utheta_{\check{q}+1}\right]$, and $\utheta_{\check{q}}$ serves as both a coarse estimate of the AoA and the initial point in the gradient-based algorithm.

\subsubsection{Multiple Paths}\label{sec:ML_ULA_multi_paths}
The algorithm can be applied to detect multiple AoAs. 
\re{When the paths are separated by at least $2\theta_{res}$ with angular resolution $\theta_{res}\approx \frac{1.78}{M_r-1}$ \cite{richards2005fundamentals}, the proposed algorithm can resolve individual paths using the same likelihood function in \eqref{eq:obj_coh_ULA}.}
\re{Assuming that the number of paths $L$ is known,} we detect $L$ AoAs by identifying the $L$ most prominent peaks from the sampled objective function,  employing the sampling policy~\eqref{eq:crude_sample}. Each peak is then refined within a constrained region using a gradient-based method to yield the final AoA estimates for all $L$ paths. This approach is capable of handling multiple paths with varying strengths.
\re{Note that the assumptions of path separation and known $L$ may not hold in realistic channels. In such cases, the algorithm can still identify the strongest beam directions.}

\subsection{ML-based Angle Estimation for Narrowband Coherent Channel with UPA and Repetitive Pilot Symbols}
\label{sec:coherent_UPA}
% When employing a uniform planar array (UPA) configuration, the antenna response is characterized by pairs of angles (\re{elevation} and azimuth). 
With a UPA, let $\theta\in\left[-\frac{\pi}{2},\frac{\pi}{2}\right]$ represent the \re{elevation} angle and $\varphi\in\left[-\frac{\pi}{2},\frac{\pi}{2}\right]$ represent the azimuth angle at the receiver.
Assume that the antenna spacing, both horizontally and vertically, is half of the wavelength. 
Let $M_{H}$ and $M_{V}$ represent the horizontal and vertical dimensions of the antenna array, respectively. Define
% \begin{align}
% &\bee_{M_H}(\varphi,\theta)\nonumber\\
% &\equiv\left[1,e^{j \pi\cos(\theta)\sin(\varphi)},\cdots,e^{j \pi\cos(\theta)\sin(\varphi)(M_H-1)}\right]^T.
% \end{align}
\begin{align}
\bee_{M_H}(\varphi,\theta)\equiv\left[1,e^{j \pi\cos(\theta)\sin(\varphi)},\cdots,e^{j \pi\cos(\theta)\sin(\varphi)(M_H-1)}\right]^T.
\end{align}
With $M = M_H \times M_V$ antennas, the antenna response vector can be expressed as
\begin{align}\label{eq:UPA_response}
\ba_M(\varphi,\theta) = \ba_{M_V}(\theta) \otimes \bee_{M_H}(\varphi,\theta),
\end{align}
where $\otimes$ denotes the Kronecker product. 
\dg{(The UPA response $\ba_M(\varphi,\theta)$ defined in~\eqref{eq:UPA_response} is not to be confused with the ULA response $\ba_M(\theta)$ defined in~\eqref{eq:ULA_reponse}.)}
% Here, we slightly abuse the notation by using $\ba(\cdot)$ to represent both ULA response and UPA response. When presented with a single argument, it refers to a ULA response, and when there are two arguments, it signifies a UPA response.
From the \re{UPA} response vector $\ba_{M_r}(\varphi,\theta)$ at the receiver, the response of the $i$-th column of the antenna array can be expressed as $\tilde{\ba}_i(\varphi,\theta)=e^{j \pi\cos(\theta)\sin(\varphi)(i-1)}\ba_{M_{V,r}}(\theta)$,
where $i=1,\cdots,M_{H,r}$. This motivates us to apply the algorithm developed for ULA case to estimate $\theta$ first. 
We \re{describe the detailed estimation solution} in the following.

\subsubsection{A single path} When there is a single path, we neglect the subscript $l$ in~\eqref{eq:flat_channel}.
We first estimate the \re{elevation} angle $\theta$ by dividing the measurements into $M_{H,r}$ groups corresponding to $M_{H,r}$ columns of the antenna array. Denote by the measurements at time $\tau$ corresponding to the $i$-th column of antenna array as
$\tilde{\br}_{i}[\tau]=\calQ(\tilde{\alpha}_i\ba_{M_{V,r}}(\theta)\ba_{M_t}^{{H}}(\omega,\psi)\bs+\bw[\tau])$,
where $\tilde{\alpha}_i=e^{j \pi\cos(\theta)\sin(\varphi)(i-1)}\alpha$. By letting $\zeta_i=\tilde{\alpha}_i\ba_{M_t}^{{H}}(\omega,\psi)\bs$, the treatment for ULA case applies. The likelihood of observing the outputs $\tilde{\bR}_i=\left[\tilde{\br}_{i}[1],\tilde{\br}_{i}[2],\cdots,\tilde{\br}_{i}[N_d]\right]$ corresponding to the $i$-th column of the antenna array, given the arrival \re{elevation} angle $\theta$, can be derived as
\begin{align}\label{eq:obj_coh_UPA_col}
\tilde{g}_i(\theta)= % \nonumber\\
&\sum_{\zeta_i\in\calZ}\prod_{m=1}^{M_{V,r}}f_{N_d}(\rho_m(\theta,\zeta_i), \tilde{\mu}_{im})f_{N_d}(\kappa_{m}(\theta,\zeta_i), \tilde{\nu}_{im}),
\end{align}
where $\tilde{\mu}_{im} = \sum_{\tau=1}^{N_d}\big(\real(\tilde{r}_{im}[\tau])+1\big)/2$ and $\tilde{\nu}_m = \sum_{\tau=1}^{N_d}\big(\imag(\tilde{r}_{im}[\tau])+1\big)/2$ with $\tilde{r}_{im}[\tau]$ being the $m$-th element of $\tilde{\mathbf{r}}_i[\tau]$.
% \begin{figure*}[b]
% \hrulefill
% \begin{align}
% \tilde{g}_i(\theta)= % \nonumber\\
% &-\log\left(\sum_{\zeta_i}\prod_{m=1}^{M_{V,r}}\Big(\erfc\big(\kappa_{m,R}(\theta,\zeta_i)\big)\Big)^{\mu_{m,R}\left(\tilde{\bR}_i\right)}\Big(1+\erf\big(\kappa_{m,R}(\theta,\zeta_i)\big)\Big)^{N_d-\mu_{m,R}\left(\tilde{\bR}_i\right)}\right.\nonumber\\
% &\left.\qquad\qquad\cdot\Big(\erfc\big(\kappa_{m,I}(\theta,\zeta_i)\big)\Big)^{\mu_{m,I}\left(\tilde{\bR}_i\right)}\Big(1+\erf\big(\kappa_{m,I}(\theta,\zeta_i)\big)\Big)^{N_d-\mu_{m,I}\left(\tilde{\bR}_i\right)}\right),\label{eq:obj_coh_UPA_col}
% \end{align}
% \begin{align}
% \bar{g}(\varphi)&=-\log\left(\sum_{\bar{\zeta}}\prod_{m=1}^{M_r}\Big(\erfc\left(\bar{\kappa}_{m,R}\left(\varphi,\bar{\zeta}\right)\right)\Big)^{\mu_{m,R}(\bR)}\Big(1+\erf\left(\bar{\kappa}_{m,R}\left(\varphi,\bar{\zeta}\right)\right)\Big)^{N_d-\mu_{m,R}(\bR)}\right.\nonumber\\
% &\left.\qquad\qquad\cdot\Big(\erfc\left(\bar{\kappa}_{m,I}\left(\varphi,\bar{\zeta}\right)\right)\Big)^{\mu_{m,I}(\bR)}\Big(1+\erf\left(\bar{\kappa}_{m,I}\left(\varphi,\bar{\zeta}\right)\right)\Big)^{N_d-\mu_{m,I}(\bR)}\right).\label{eq:obj_coh_varphi}
% \end{align}
% \end{figure*}
Combining the \re{likelihood functions} across all antenna array columns involves handling distinct $\zeta_i$ values. While these values are correlated, we assume independence \re{for tractability}, leading to the following objective function
\begin{align}\label{eq:obj_coh_theta}
\tilde{g}(\theta)=\prod_{i=1}^{M_{H,r}}\tilde{g}_i(\theta).
\end{align}
We solve the problem $\max_{\theta}\tilde{g}(\theta)$ using a two-step estimation as outlined in Sec. \ref{sec:ML_ULA_single_path}.
After obtaining the estimate $\hat{\theta}$, we substitute it into $\ba_{M_r}(\varphi,\theta)$ in \eqref{eq:UPA_response}. 
Define
\begin{align}
&\bar{\rho}_m(\varphi,\theta,\zeta)\equiv\real\left(\zeta a_m(\varphi,\theta)\right),\\
&\bar{\kappa}_m(\varphi,\theta,\zeta)\equiv\imag\left(\zeta  a_m(\varphi,\theta)\right),
\end{align}
where $a_m(\varphi,\theta)$ is the $m$-th element of the antenna response $\ba_M(\varphi,\theta)$.
Letting $\bar{\zeta}=\alpha\mathbf{a}_{M_t}^H(\omega,\psi)\bs$, the objective function with respect to $\varphi$ can be expressed as
\begin{align}\label{eq:varphi_coherent}
\bar{g}(\varphi)&=\sum_{\bar{\zeta}\in\calZ}\prod_{m=1}^{M_r}
f_{N_d}\left(\bar{\rho}_m\left(\varphi,\hat{\theta},\bar{\zeta}\right), \mu_{m}\right)\nonumber\\
&\qquad\qquad\qquad\qquad\cdot f_{N_d}\left(\bar{\kappa}_{m}\left(\varphi,\hat{\theta},\bar{\zeta}\right), \nu_{m}\right).
\end{align}
We then estimate $\hat{\varphi}$ using the same two-step approach.
% In particular, let $\bar{\zeta}=\alpha\ba_{M_t}^{{H}}(\omega,\psi)\bs$, $\bar{\kappa}_{m,R}\left(\varphi,\bar{\zeta}\right) = -\real\left(\bar{\zeta}a_m\left(\varphi,\hat{\theta}\right)\right)$ and $\bar{\kappa}_{m,I}\left(\varphi,\bar{\zeta}\right) = -\imag\left(\bar{\zeta}a_m\left(\varphi,\hat{\theta}\right)\right)$. With these changes, we obtain the objective function~\eqref{eq:obj_coh_varphi} to solve $\ba_{M_{V,r}}_{\varphi}\quad\bar{g}(\varphi)$. 
% Therefore, we obtain both angle estimations $\hat{\theta}$ and $\hat{\varphi}$.

\subsubsection{Multiple paths} 
\re{When the paths are separated by at least $2\theta_{res}\approx\frac{3.56}{M_{V,r}-1}$ in the \re{elevation} angle direction and $2\varphi_{res}\approx\frac{3.56}{M_{H,r}-1}$ in the azimuth angle direction, the proposed algorithm can resolve individual paths. Assuming the number of paths $L$ is known a priori, we} first obtain fine estimates for the $L$ \re{elevation} angles, denoted as $\hat{\theta}_1,\cdots,\hat{\theta}_L$, using the objective function in~\eqref{eq:obj_coh_theta}. The correspondence between each \re{elevation} and azimuth angle pair is established through sequential estimation. Specifically, to estimate $\varphi_l$, we substitute $\hat{\theta}_l$ into the antenna response and obtain the associated objective function as expressed in ~\eqref{eq:varphi_coherent}. The solution $\hat{\varphi}_l$, together with $\hat{\theta}_l$, forms the angle pair estimation for path $l$.

\subsection{Adaptations to Wideband Coherent Channel}\label{sec:wideband_ext}
\re{In this subsection, we first examine the adaptation of the algorithm to the narrowband noncoherent channel, defined as
\begin{align}\label{eq:nonco_channel}
\bH[\tau]=\sum_{l=1}^L\alpha_l[\tau]\ba_{M_r}(\theta_l,\varphi_l)\ba_{M_t}^{{H}}(\psi_l,\omega_l),
\end{align}
using a UPA as an example. In this model, path coefficients $\alpha_l[\tau]$'s are independent across $\tau$, while path angles remain constant. 
Subsequently, we demonstrate how the treatment to this model can be applied to channel estimation in wideband coherent scenarios as described in Eq. \eqref{eq:Cluster_channel}.} 

We assume that the transmitted pilot sequence $\mathbf{S}=[\mathbf{s}[1],\mathbf{s}[2],\ldots,\mathbf{s}[N_d]] \in \mathbb{C}^{M_t \times N_d}$ is unknown and lacks a discernible pattern.
We sequentially estimate $\theta$ and then $\varphi$, deriving new likelihood-based objective functions to address channel non-coherence and varying transmitted pilots. Define $\tilde{\mu}_{im}^{\tau}=(\real(\tilde{r}_{im}[\tau])+1)/2$ and $\tilde{\nu}_{im}^{\tau}=(\imag(\tilde{r}_{im}[\tau])+1)/2$. The likelihood function given $\theta$ can be expressed as
% \begin{align}\label{eq:like_nonco_theta}
% \tilde{h}(\theta)=\prod_{i=1}^{M_{H,r}}\tilde{h}_i(\theta),
% \end{align}
% where 
% \begin{align}\label{eq:obj_nonco_theta}
% &\tilde{h}_i(\theta)\nonumber\\
% &=\prod_{\tau=1}^{N_d}\left(\sum_{\zeta\in\calZ}\left(\prod_{m=1}^{M_{V,r}}f_{1}(\rho_m(\theta,\zeta),\tilde{\mu}_{im}^{\tau})f_{1}(\kappa_m(\theta,\zeta),\tilde{\nu}_{im}^{\tau})\right)\right)
% \end{align} 
% is the likelihood of observing $\tilde{\bR}_i$ given $\theta$.
\begin{align}\label{eq:like_nonco_theta}
\tilde{h}(\theta)&=\prod_{i=1}^{M_{H,r}}\prod_{\tau=1}^{N_d}\Bigg(\sum_{\zeta\in\calZ}\prod_{m=1}^{M_{V,r}}f_{1}(\rho_m(\theta,\zeta),\tilde{\mu}_{im}^{\tau})\nonumber\\
&\qquad\qquad\qquad\qquad\cdot f_{1}(\kappa_m(\theta,\zeta),\tilde{\nu}_{im}^{\tau})\Bigg).
\end{align}
Compared to~\eqref{eq:obj_coh_theta}, the objective function in~\eqref{eq:like_nonco_theta} combines the measurements across time slots in a different way to take into account the varying path gain $\zeta[\tau]=\alpha[\tau]\ba_{M_t}^{{H}}(\psi,\omega)\bs[\tau]$. 
With an estimate $\hat{\theta}$, we can obtain the corresponding $\varphi$ estimate with the objective function 
\begin{align}\label{eq:like_nonco_ULA}
\bar{h}(\varphi)&=\prod_{\tau=1}^{N_d}\Bigg(\sum_{\bar{\zeta}\in\calZ}\prod_{m=1}^{M_r}f_{1}\left(\bar{\rho}_m\left(\varphi,\hat{\theta},\bar{\zeta}\right),\mu_m^{\tau}\right)\nonumber\\
&\qquad\qquad\qquad\qquad\cdot f_{1}\left(\bar{\kappa}_m\left(\varphi,\hat{\theta},\bar{\zeta}\right),\nu_m^{\tau}\right)\Bigg),
\end{align}
where $\mu_m^{\tau}=\big(\real(r_m[\tau])+1\big)/2,\nu_m^{\tau}=\big(\imag(r_m[\tau])+1\big)/2$.

We can leverage the narrowband noncoherent channel model to address the angle estimation problem in wideband coherent channels characterized by unknown delay spreads and filtering effects. In particular, with channel model~\eqref{eq:Cluster_channel}, the term $\alpha_l \sum_{d=0}^{D-1} \bs[\tau-d] p(dT - \delta_l)$ can be treated as the unknown and time-varying $\alpha_l[\tau] \bs[\tau]$ in \eqref{eq:nonco_channel}, enabling the application of the same methodology.

% Utilizing the MLE for the noncoherent channel, however, introduces increased computational complexity. The reason can be briefly explained as the following. At time slot $\tau$, the likelihood of observing $\br[\tau]$ given the arrival angle $\theta$ and path information $\zeta[\tau]$ is
% \begin{align}
% f(\br[\tau]|\theta,\zeta[\tau]).
% \end{align}
% In the coherent case, $\zeta[\tau]$ is a constant for all $\tau$. We can obtain the likelihood regarding the received signal $\bR$ as
% \begin{align}
% f(\bR|\theta)\propto\sum_{\zeta}f(\bR|\theta,\zeta)=\sum_{\zeta}\prod_{\tau=1}^{N_d}f(\br[\tau]|\theta,\zeta),
% \end{align}
% where we only need to address the unknown $\zeta$ once, after we combine over all time slots.
% Suppose unknown $\zeta[\tau]$ varies over time slots. We need to deal with the unknown $\zeta[\tau]$ before combing over time slots, the likelihood thus follows as
% \begin{align}
% f(\bR|\theta)=\prod_{\tau=1}^{N_d}f(\br[\tau]|\theta)\propto\prod_{\tau=1}^{N_d}\left(\sum_{\zeta[\tau]}f(\br[\tau]|\theta,\zeta[\tau])\right).
% \end{align}
% In each time slot, the computational complexity grows with the increasing size of potential $\zeta[\tau]$ values, contributing to the overall computational load.

% We propose the non-coherent channel model and corresponding estimation methods not only to address extreme non-coherence but also to introduce additional flexibility not accounted for in the discussed coherent channel model.

\section{Analog Beamforming Design and Performance Evaluation Metrics}\label{sec:beamforming}
We aim to optimize analog beamforming by leveraging angular-domain channel estimation to maximize the post-beamforming SNR. Our design focuses on coherent channel scenarios, illustrated using a UPA.

\subsection{Analog Beamforming for Narrowband Coherent Channels with Repetitive Pilot Symbols}
Assuming $\bs[\tau] = \bs$, we examine various beamforming schemes. For the channel in~\eqref{eq:flat_channel}, 
problem \ref{eq:beamforming_prob} reduces to
\begin{align}
\max_{\bb:|b_1|=\cdots=|b_{M_r}|=1}\quad{\left\vert\bb^{{H}}\left(\sum\nolimits_{l=1}^L\zeta_l\ba_{M_r}(\varphi_l,\theta_l)\right)\right\vert^2}.
\end{align}
\subsubsection{Ideal beamforming} Given perfect knowledge of $\zeta_l,\varphi_l$ and $\theta_l$, the ideal beamforming vector can be derived as
\begin{align}
\bb_{IDEAL}=\exp\left(j \angle\left(\sum\nolimits_{l=1}^L\zeta_l\ba_{M_r}(\varphi_l,\theta_l)\right)\right),
\end{align}
where $\angle(\cdot)$ extracts the element-wise phases, and $\exp(\cdot)$ computes the element-wise exponentials. 

\subsubsection{Estimation beamforming} 
After estimating AoAs, we apply ML estimation to obtain estimates of $\zeta_l$. The resulting beamforming can be represented as 
\begin{align}
\bb_{EST}=\exp\left(j \angle\left(\sum\nolimits_{l=1}^L\hat{\zeta}_l\ba_{M_r}\left(\hat{\varphi}_l,\hat{\theta}_l\right)\right)\right)
\end{align}
% \begin{figure*}[b]
% \hrulefill
% \begin{align}
% \underline{g}(\beta)&=-\log\left(\prod_{m=1}^{M_r}\Big(\erfc\left(\underline{\kappa}_{m,R}\left({0.1,\beta}\right)\right)\Big)^{\mu_{m,R}(\bR)}\Big(1+\erf\left(\underline{\kappa}_{m,R}\left({0.1,\beta}\right)\right)\Big)^{N_d-\mu_{m,R}(\bR)}\right.\nonumber\\
% &\left.\qquad\qquad\cdot\Big(\erfc\left(\underline{\kappa}_{m,I}\left({0.1,\beta}\right)\right)\Big)^{\mu_{m,I}(\bR)}\Big(1+\erf\left(\underline{\kappa}_{m,I}\left({0.1,\beta}\right)\right)\Big)^{N_d-\mu_{m,I}(\bR)}\right).\label{eq:obj_coh_zeta}
% \end{align}
% \begin{align}
% \check{g}(\bgamma)&=-\log\left(\prod_{m=1}^{M_r}\left(\erfc\left(\sum_{l=1}^L\underline{\kappa}_{m,R}\left(\gamma_l,\hat{\beta}_l\right)\right)\right)^{\mu_{m,R}(\bR)}\left(1+\erf\left(\sum_{l=1}^L\underline{\kappa}_{m,R}\left(\gamma_l,\hat{\beta}_l\right)\right)\right)^{N_d-\mu_{m,R}(\bR)}\right.\nonumber\\
% &\left.\qquad\qquad\cdot\left(\erfc\left(\sum_{l=1}^L\underline{\kappa}_{m,I}\left(\gamma_l,\hat{\beta}_l\right)\right)\right)^{\mu_{m,I}(\bR)}\left(1+\erf\left(\sum_{l=1}^L\underline{\kappa}_{m,I}\left(\gamma_l,\hat{\beta}_l\right)\right)\right)^{N_d-\mu_{m,I}(\bR)}\right)\label{eq:obj_coh_gamma}
% \end{align}
% \end{figure*}

\subsubsection{Strong beamforming} 
For the $l$-th estimated path angle pair $(\hat{\varphi}_l, \hat{\theta}_l)$, we test the beamformer $\bb_l = \ba_{M_r}(\hat{\varphi}_l, \hat{\theta}_l)$ by applying it to the quantized received signals and calculating the average received energy $\overline{\vert y_l[\tau]\vert^2}$ across $N_d$ time slots. We then select the beamformer as
\begin{align}
\bb_{STR}=\bb_l \quad\text{such that}\quad l=\newargmax\overline{{\vert y_l[\tau]\vert^2}}.
\end{align}

The SNR associated with \textit{Ideal beamforming} serves as the upper limit for all practical beamforming designs. We define the average SNR ratio (associated with beamformer $\bb$) compared to the ideal beamforming as
\begin{align}
\eta&=\frac{1}{K}\sum\nolimits_{k=1}^K \left(\left\vert\bb^{H}\bx^{(k)}\right\vert^2/{\left\vert(\bb_{IDEAL})^{H}\bx^{(k)}\right\vert^2}\right)\label{eq:SNR_ratio}
\end{align}
where $k$ indexes the independent random realizations, and $K$ represents the total number of conducted realizations.

\subsection{Analog Beamforming for Wideband Coherent Channels with Unknown Pilot Sequence}\label{sec:wideband_beamforming}
Given unknown $\bS=[\bs[1],\cdots,\bs[N_d]]$ and wideband channels, we redefine the beamforming design.
\subsubsection{Wideband optimal beamforming} The wideband ideal beamforming vector is found through problem \eqref{eq:beamforming_prob}.
% the following problem
% \begin{align}\label{eq:wide_ideal_beamforming}
% \max_{\bb:\vert b_m\vert=1,\forall m=1,\cdots,M_r}\quad{\bb^{H}\bbE\left[\bz_{\text{sig}}[\tau]\bz_{\text{sig}}[\tau]^{H}\right]\bb}
% \end{align}
Suppose $\bx[\tau]$ is observable. We approximate $\bbE_{\tau}\left[\bx[\tau]\bx[\tau]^{H}\right]$ as $\bar{\bX}=\sum_{\tau=1}^{N_d}\bx[\tau]\bx[\tau]^{H}/N_d$. The optimal beamformer is defined as
\begin{align}
\bb_{WOPT}=\newargmax_{\bb:|b_1|=\cdots=|b_{M_r}|=1}\quad\bb^{{H}}\bar{\bX}\bb.
\end{align}
% Given the constraint $\vert\bb\vert^2=1$, $\bb^{H}\bar{\bX}\bb$ is maximized by choosing $\bb$ as the eigenvector corresponding to the largest eigenvalue of $\bar{\bX}$ \cite{allaire2008numerical}, as $\bar{\bX}$ is Hermitian. 
% Such $\bb$ can be found through SVD of $\bar{\bZ}_{\text{sig}}$ and choose either the right or left singular vector corresponding to the maximum singular value of $\bz_{\text{sig}}$. 
\re{Numerically, the suboptimal solution can be obtained using a block coordinate descent algorithm to optimize the phases $\boldsymbol{\xi}=[\xi_1,\cdots,\xi_{M_r}]^T$  with an initial point $\angle(\bb^0)$, where $\bb^0$ denotes the normalized eigenvector associated with the largest eigenvalue of $\bar{\bX}$.}

\subsubsection{Wideband unquantized beamforming} Suppose $\bz[\tau]$ is observable.
% Note that $\bbE_{\tau}\left[\left\vert\bb^{{H}}\bx[\tau]\right\vert^2\right]
% =\bb^{H}\bbE_{\tau}\left[\bz[\tau]\bz[\tau]^{H}\right]\bb - M_r$.
% \begin{align}
% \bbE\left[\left\vert\bb^{{H}}\bz_{\text{sig}}[\tau]\right\vert^2\right]&=
% \bbE\left[\left\vert\bb^{{H}}(\bz[\tau]-\bw[\tau]))\right\vert^2\right]\\
% % &=\bb^{H}\bbE\left[\bz[\tau]\bz[\tau]^{H}\right]\bb - \bb^{H}\bbE\left[\bw[\tau]\bw[\tau]^{H}\right]\bb \\
% &=\bb^{H}\bbE\left[\bz[\tau]\bz[\tau]^{H}\right]\bb - M_r.
% \end{align}
The objective in problem \eqref{eq:beamforming_prob} can be equivalently written as $\bb^{H}\bbE_{\tau}\left[\bz[\tau]\bz[\tau]^{H}\right]\bb$. 
We approximate $\bbE_{\tau}\left[\bz[\tau]\bz[\tau]^{H}\right]$ as $\bar{\bZ}=\sum_{\tau=1}^{N_d}\bz[\tau]\bz[\tau]^{H}/N_d$. The beamformer is then determined by
\begin{align}
\bb_{WUNQ}=\newargmax_{\bb:\vert b_1\vert=|b_2|=\cdots=|b_{M_r}|=1}\quad\bb^{{H}}\bar{\bZ}\bb.
\end{align}

\subsubsection{Wideband quantized beamforming} With 1-bit ADCs, we can only observe the quantized signal in the form of~\eqref{eq:r[tau]}.
\re{If} the quantized signal $\br[\tau]$ retains a significant amount of information of $\bx[\tau]$, the beamforming vector $\mathbf{b}$ that maximizes $\bb^{H}\bbE_{\tau}\left[\br[\tau]\br[\tau]^{H}\right]\bb$ should yield a reasonably effective beamforming design. We thus find the beamformer as
\begin{align}
\bb_{WQ}=\newargmax_{\bb:|b_1|=\cdots=|b_{M_r}|=1}\quad\bb^H\bar{\bR}\bb,
\end{align}
whith $\bar{\bR}=\sum_{\tau=1}^{N_d}{\br[\tau]\br[\tau]^H}/{N_d}$.

\subsubsection{Wideband strong beamforming} 
By employing the treatment outlined in Section \ref{sec:wideband_ext}, we obtain the angular estimate $\left(\hat{\varphi},\hat{\theta}\right)$ by setting $L=1$, and choose the beamforming as 
\begin{align}
\bb_{WSTR}=\ba_{M_r}\left(\hat{\varphi},\hat{\theta}\right).
\end{align}

% We evaluate the performance of different beamforming schemes within the CDL-C channel model framework, characterized by clustered paths sharing similar arrival angles. This clustering poses challenges for accurate angle estimation because of the collective influence of nearby paths. 
\textit{Wideband optimal beamforming} serves as an upper bound for SNR evaluation. While statistically optimal, \textit{Wideband unquantized beamforming} may degrade in performance due to random noise patterns. Unlike these two designs which depend on unquantized signals, \textit{Wideband quantized beamforming} and \textit{Wideband strong beamforming} are \re{practical schemes given 1-bit ADCs}. Our objective is to demonstrate that, by utilizing our angular-domain estimation treatment, \textit{Wideband strong beamforming} effectively captures the dominant energy from all paths, highlighting its practical utility.

\section{Simulation Results}\label{sec:simulation}
\subsection{Narrowband Coherent Channel}
We generate AoDs/AoAs uniformly in \(\left[-\pi/3, \pi/3\right]\), subject to the \re{aforementioned} angular separation. We sample $\bs$ from $\calCN(\mathbf{0}_{M_t}, \mathbf{I}_{M_t})$ and define $\bS$ as repetitions of $\bs$. Results are averaged over 500 independent realizations. 

\begin{figure}[htbp]
\centering
\includegraphics[width=0.3\textwidth]{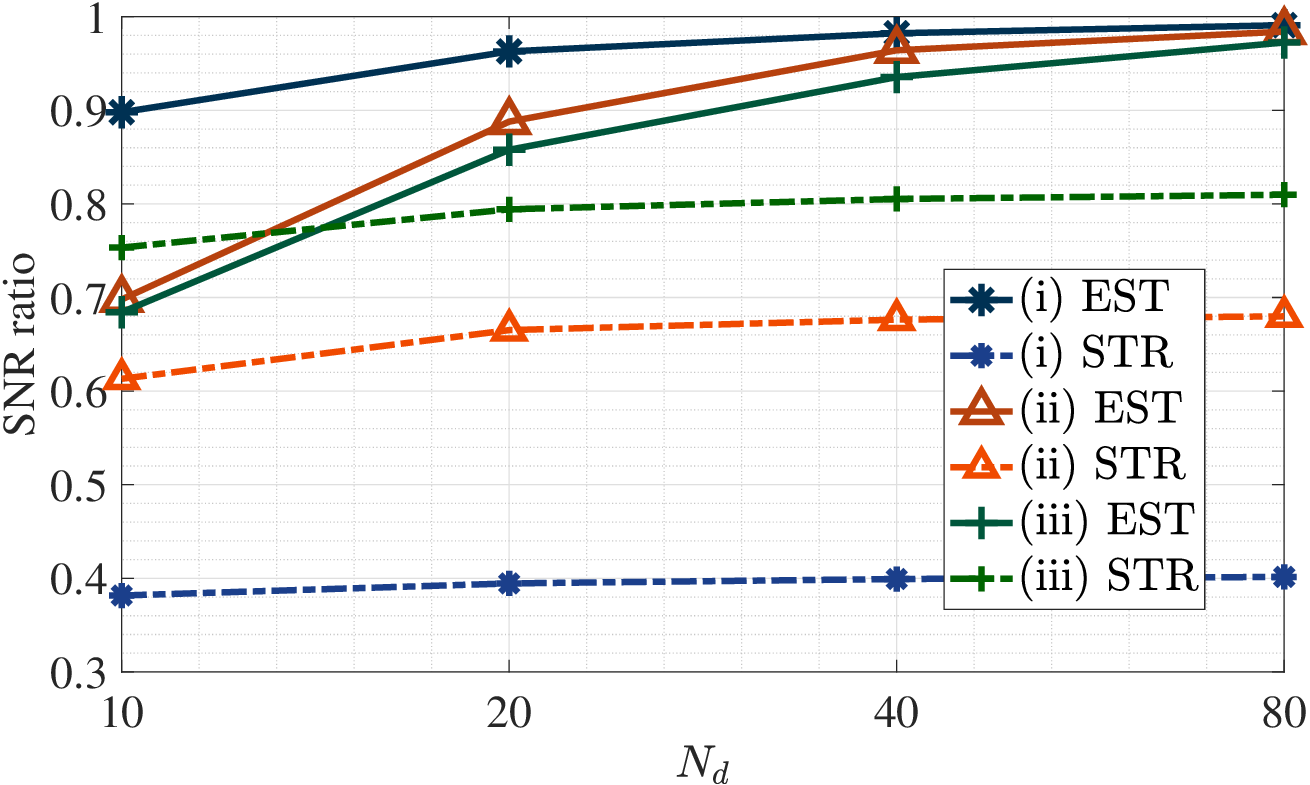}
\caption{Average SNR ratio after beamforming in a narrowband coherent channel with $L=3$. Our proposed \textit{Estimation beamforming} (EST) recovers over 90\% of the received power compared to beamforming with full channel knowledge under various path SNRs. \textit{Strong beamforming} (STR), focusing on a specific direction, is suboptimal in this case.}
\label{fig:snr_coherent_upa_lp3_mr256}
\end{figure}

We present beamforming results for multiple paths with \(L=3\), representing the limited number of main clusters typical in mmWave communication. Using UPAs with \(M_t=4 \times 4\) and \(M_r=16 \times 16\), performance is evaluated by varying the SNR and pilot length \(N_d\). 
The three SNR configurations considered are: (i) equal \re{path} SNRs of $-18$ dB for all paths, (ii) SNRs of $-18$, $-21$, and $-24$ dB, and (iii) SNRs of $-18$, $-23$, and $-28$ dB. The average ideal SNR after beamforming for these cases is approximately $10$ dB, $7.7$ dB, and $7$ dB, respectively. The SNR performance is shown in Fig.~\ref{fig:snr_coherent_upa_lp3_mr256}.
While not depicted, the average angle estimation error for eight of the nine examined paths is below $0.1$ radians for both \re{elevation} and azimuth angles with $N_d=40$, except for the path with an SNR of $-28$ dB. \re{With accurate AoA estimations, the SNR ratio exceeds $0.9$ across all three SNR settings when applying the proposed \textit{Estimation beamforming}, compared to the ideal beamforming case, provided $N_d\geq 40$.}

\subsection{CDL-C Channel Model}
We evaluate the proposed analog beamforming in wideband coherent channels using the CDL-C model, characterized by clustered paths sharing similar departure/arrival angles. Despite challenges from clustered features and azimuth angles spanning \(\varphi \in [-\pi, \pi]\), our focus is on identifying a single beam direction that captures significant energy from all paths.
We fix \( M_t = 1 \) and \( M_r = 16 \times 16 \), and use OFDM modulation with a carrier frequency of 28 GHz and subcarrier spacing of 240 kHz.  

\begin{figure}[htbp]
\centering
\includegraphics[width=0.3\textwidth]{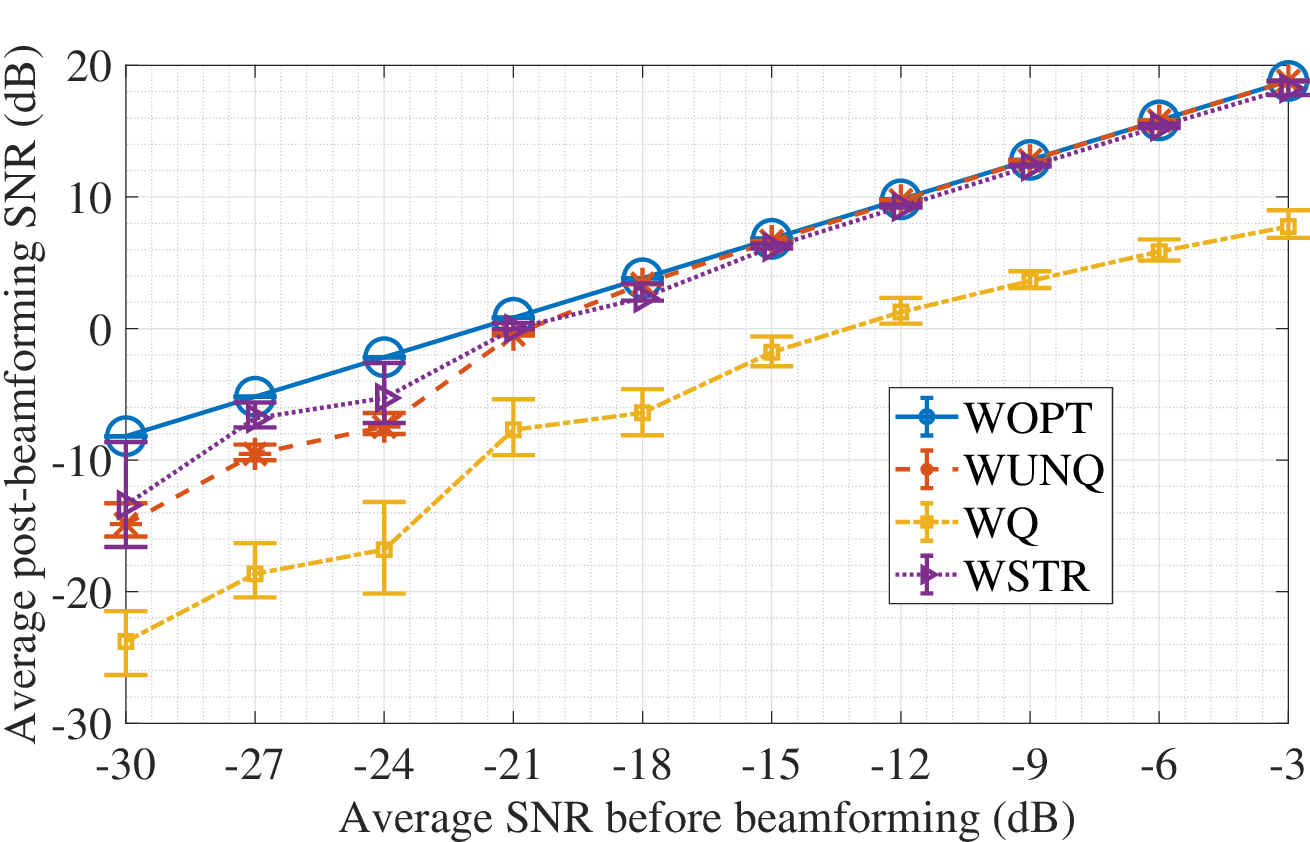}
\caption{Average SNR after beamforming under the CDL-C model. Markers represent mean values across 60 realizations, with upper/lower error bars corresponding to the upper and lower 25\% quantiles. \textit{Wideband strong beamforming} (WSTR) closely approximates optimal performance across SNRs from \(-1\) dB to 20 dB.}
\label{fig:snr_wideband}
\end{figure}

The evaluation varies SNRs before analog beamforming to assess average post-beamforming SNR.
The results are shown in Fig. \ref{fig:snr_wideband}.
For \textit{Wideband strong beamforming} and \textit{Wideband quantized beamforming}, we use $N_d = \{12288, 4096, 1024, 512, 128\}$ for SNRs $\{-30, -27, -24, -21, -18\}$ dB before beamforming, and halve $N_d$ for each 3 dB increase afterwards, reaching $N_d = 4$ at $-3$ dB. For comparison, pilot lengths for \textit{Wideband optimal/unquantized beamforming} are set to at least 512 if $N_d$ chosen for \textit{Wideband strong/quantized beamforming} is smaller than that.
\re{While longer pilots are required for very low SNRs, in the post-beamforming SNR range of around 0 dB and above, \textit{Wideband strong beamforming} achieves near-optimal performance with around 1000 or much fewer pilot chips. Since an OFDM symbol typically includes a few thousand chips, beam acquisition can be completed using a one-shot pilot transmission within a single OFDM symbol in this SNR region, thus significantly reducing latency and overhead.}

Discrepancies arise between \textit{Wideband optimal beamforming} and \textit{Wideband unquantized beamforming} at low SNRs, where noise power dominates. In such cases, \textit{Wideband strong beamforming} can even outperform \textit{Wideband unquantized beamforming}. 
Additionally, \textit{Wideband quantized beamforming} appears suboptimal due to quantization losses that hinder effective information extraction.
In contrast, \textit{Wideband strong beamforming}, utilizing quantized signals with limited pilot length, demonstrates robust performance across a wide SNR range ($-1$ dB to $20$ dB post-beamforming), closely approximating \textit{Wideband optimal beamforming}. This underscores the practicality and effectiveness of the proposed scheme.

\section{Conclusion}
We introduced angular-domain channel estimation and analog beamforming design, leveraging additional hardware in the form of full digital \re{receive} chains equipped with 1-bit ADCs. Our methods were evaluated in narrowband coherent channels, with the analysis focusing on scenarios where each path maintains a separation of at least $2\theta_{res}$ (and $2\varphi_{res}$) from all other paths, and where the number of paths $L$ is known in advance. We plan to extend this method to address these restrictions in a more comprehensive future paper. 
In these scenarios, accurate angular estimations allow our proposed \textit{Estimation beamforming} to effectively incorporate all estimated paths into a joint beamforming design, thereby maximizing energy capture. Additionally, we tested our methods in wideband coherent channels under the CDL-C channel model, characterized by more complex clustered conditions. Rather than precisely identifying each cluster, we advocate for \textit{Wideband strong beamforming} to determine a beam direction that captures significant energy from all clusters. 
Simulation results demonstrate the efficacy of our proposed schemes in improving \re{post-beamforming SNR and hence the feasibility of one-shot beam acquisition based on the newly proposed receive architecture with minimal additional cost and complexity}, highlighting the potential utility for commercial mmWave MIMO communication systems.

\bibliographystyle{IEEEtran}
% \bibliography{IEEEabrv,def,dguo,all}
\bibliography{IEEEabrv,ref}

\end{document}